\documentclass{article}%
\usepackage{graphicx}
\usepackage{amsmath}
\usepackage{amsfonts}
\usepackage{amssymb}%
\begin{document}

\author{Antony Valentini\\Augustus College}

\begin{center}
{\LARGE Hidden Variables and the Large-Scale Structure of Spacetime}

\bigskip\bigskip

\bigskip Antony Valentini

\bigskip

\bigskip

\textit{Perimeter Institute for Theoretical Physics,}

\textit{31 Caroline Street North, Waterloo, Ontario N2L 2Y5, Canada.}

(e-mail: avalentini@perimeterinstitute.ca)
\end{center}

\bigskip

We discuss how to embed quantum nonlocality in an approximately classical
spacetime background, a question which must be answered irrespective of any
underlying microscopic theory of spacetime. We argue that, in deterministic
hidden-variables theories, the choice of spacetime kinematics should be
dictated by the properties of generic non-equilibrium states, which allow
nonlocal signalling. Such signalling provides an operational definition of
absolute simultaneity, which may naturally be associated with a preferred
foliation of classical spacetime. The argument applies to any deterministic
hidden-variables theory, and to both flat and curved spacetime backgrounds. We
include some critical discussion of Einstein's 1905 `operational' approach to
relativity, and compare it with that of Poincar\'{e}.

\bigskip

\bigskip

\textbf{CONTENTS}

\ 

\textbf{1 Introduction}

-- 1.1 Status of Lorentz Invariance in Contemporary Physics

-- 1.2 Quantum Nonlocality

-- 1.3 Non-Equilibrium Hidden Variables

\textbf{2 Physical Structure of Spacetime}

-- 2.1 Kinematics and Dynamics

-- 2.2 Einstein and Poincar\'{e} in 1905

\textbf{3 Instantaneous Signalling in Quantum Non-Equilibrium}

-- 3.1 General (Deterministic) Hidden-Variables Theories

-- 3.2 The Example of Pilot-Wave Theory

\textbf{4 Absolute Simultaneity in Flat and Curved Spacetime}

-- 4.1 Flat Spacetime

-- 4.2 Curved Spacetime

\textbf{5 Discussion and Conclusion}

\bigskip

To be published in: \textit{Absolute Simultaneity}, eds. W. L. Craig and Q.
Smith (Routledge, London, 2005).

\bigskip

\bigskip

\bigskip

\bigskip

\bigskip

\bigskip

\bigskip

\bigskip

\bigskip

\bigskip

\bigskip

\bigskip

\bigskip

\bigskip

\bigskip

\bigskip

\bigskip

\bigskip

\bigskip

\bigskip

\bigskip

\bigskip

\bigskip

\bigskip

\bigskip

\bigskip

\bigskip

\bigskip

\bigskip

\bigskip

\bigskip

\bigskip

\bigskip

\bigskip

\bigskip

\bigskip

\bigskip

\bigskip

\bigskip

\bigskip

\bigskip

\bigskip

\bigskip

\bigskip

\bigskip

\bigskip

\bigskip

\bigskip

\bigskip

\bigskip

\bigskip

\bigskip

\bigskip

\bigskip

\bigskip

\bigskip

\bigskip

\bigskip

\bigskip

\bigskip

\bigskip

\bigskip

\bigskip

\bigskip

\bigskip

\bigskip

\bigskip

\bigskip

\bigskip

\bigskip

\bigskip

\bigskip

\bigskip

\section{Introduction}

\begin{quotation}
`\textit{The simultaneity of two events, or the order of their succession, the
equality of two durations, are to be so defined that the enunciation of the
natural laws may be as simple as possible.}'

-- Henri Poincar\'{e} (1905a).

\ 

`\textit{What really matters is not merely the greatest possible simplicity of
the geometry alone, but rather the greatest possible simplicity of all of
physics (inclusive of geometry).}'

-- Albert Einstein (1949a).
\end{quotation}

This article concerns the structure of spacetime on large scales, in the
context of hidden-variables interpretations of quantum theory. In particular,
we shall be addressing the question of how macroscopic quantum nonlocality can
be embedded in an approximately classical spacetime background. We argue that
this question must have an answer, regardless of what the underlying
microscopic theory of spacetime may turn out to be, and further, that the most
natural answer is to introduce an absolute simultaneity, associated with a
preferred foliation of classical spacetime (flat and curved).

The introduction of an absolute simultaneity, to accommodate the nonlocality
of quantum theory over macroscopic distances, was suggested in particular by
Popper (1982), Bohm and Hiley (1984), and Bell (1986, 1987). This proposal is
often regarded as unsatisfactory, because quantum nonlocality cannot in fact
be used for practical signalling at a distance, making the preferred rest
frame undetectable in practice. As Bell put it (1986: 50):

\begin{quotation}
`It is as if there is some kind of conspiracy, that something is going on
behind the scenes which is not allowed to appear on the scenes. And I agree
that that's extremely uncomfortable' (Bell 1986).
\end{quotation}

However, Bell missed the point that, in (deterministic) hidden-variables
theories, the inability to use quantum nonlocality for remote signalling is
not a fundamental constraint. It is, rather, a peculiarity of a special
`quantum equilibrium' distribution of hidden variables. For more general,
`non-equilibrium' distributions, practical nonlocal signalling is indeed
possible (Valentini 1991a,b, 1992, 1996, 2001, 2002a,b,c). From this
perspective, our inability to detect the preferred rest frame is \textit{not}
an uncomfortable conspiracy seemingly built into the laws of physics; it is
simply an accident of our living in a state of quantum equilibrium, whose
statistical noise masks the underlying nonlocality.

In our view, if one wishes to appraise the structure of spacetime at a more
fundamental level, then this should be done taking into account the wider,
explicitly nonlocal physics of quantum non-equilibrium, rather than merely in
terms of the statistical predictions of quantum theory, which are not
fundamental but merely contingent on a special distribution of hidden variables.

We shall argue that non-equilibrium instantaneous signalling defines an
absolute simultaneity, within the approximately classical spacetime defined by
macroscopic rods and clocks, and that fundamental local Lorentz invariance
should be abandoned. It will also be suggested that the widespread excessive
reluctance to consider abandoning (local) Minkowski spacetime has its origin
in the unfortunate `operational' approach to relativity taken by Einstein in 1905.

\subsection{Status of Lorentz Invariance in Contemporary Physics}

Locally speaking, the relativity of simultaneity is usually regarded as a
consequence of (local) Lorentz invariance. Before considering quantum
nonlocality, then, let us first briefly review the current status of Lorentz
invariance in other areas of physics.

In high-energy physics, the status of Lorentz invariance is certainly open to
question. The divergences of quantum field theory can be most
straightforwardly eliminated by introducing a short-distance cutoff, which
breaks Lorentz invariance. This suggests that it would be an advantage if
Lorentz invariance were not fundamental. It is also sometimes argued that
exact Lorentz invariance is experimentally inaccessible because the boost
parameter (for the non-compact Lorentz group) has an infinite range which can
never be probed uniformly (Jacobson and Mattingly 2001).

However, like renormalisability, Lorentz invariance did play a key historical
role in the development of the standard model of particle physics. Yet, the
fact that only renormalisable terms appear in the Lagrangian of the standard
model is now generally regarded as merely an accident of the low-energy limit,
where non-renormalisable terms are screened off in the infra-red (Weinberg
1995: 519). Clearly, the mere fact that a property played a crucial historical
role in constructing our current theories is not a conclusive argument for
that property to be fundamental.

Possibly, in high-energy physics, Lorentz invariance will eventually acquire a
similar status to that of renormalisability, as a mere low-energy symmetry
(Nielsen and Ninomiya 1978; Chadha and Nielsen 1983; Allen 1997; Moffat 2003).
In any case, non-Lorentz-invariant extensions of the standard model have been
considered in detail (Colladay and Kosteleck\'{y} 1998; Coleman and Glashow
1999), where terms in the Lagrangian breaking Lorentz symmetry might come from
deeper physics beyond the standard model. A number of experiments searching
for such effects have been performed, while further experiments are underway
or being planned (for reviews, see Kosteleck\'{y} 2002).

Further questioning of Lorentz invariance comes from quantum gravity, in which
the possibility of a minimum length at the Planck scale suggests that Lorentz
invariance might emerge only as an approximation on larger scales
(Kosteleck\'{y} 2002). Indeed, it has been suggested that peculiarities in
cosmic-ray data, together with other astrophysical anomalies, might be a sign
of a breakdown of standard special-relativistic kinematics, possibly due to
quantum gravity effects (Amelino-Camelia 2002). In addition, models of
classical gravitation with a `dynamical preferred frame' have been considered
(Jacobson and Mattingly 2001).

One could certainly question the above motivations for considering a breakdown
of local Lorentz invariance. Still, it is clear that Lorentz invariance is far
from being a dogma in the context of high-energy physics or quantum gravity.
Rather, it is often regarded as one important symmetry among others, whose
status (approximate or fundamental) is a matter for experiment. And in
comparing experiments with theory, it is helpful to have models incorporating
violations of Lorentz invariance (as well as models incorporating violations
of other important symmetries such as CPT invariance; Mavromatos 2004).

\subsection{Quantum Nonlocality}

In contrast, in the context of quantum foundations, attachment to Lorentz
invariance tends to be more dogmatic. A number of authors insist that a
realistic quantum physics should be `seriously Lorentz invariant', in the
sense that Lorentz invariance should be fundamental, and not merely
phenomenological or emerging in some limit. This contrast is remarkable,
because it is precisely in quantum foundations that there is arguably the
strongest motivation of all for abandoning fundamental Lorentz invariance: the
experimental detection of quantum nonlocality, through the observed violations
of Bell's inequalities.

As emphasised by Bell (1987), quantum theory is incompatible with locality,
\textit{independently} of any assumption about the existence of hidden
variables. Given a pair of spin-1/2 particles in the singlet state, a quantum
measurement of $z$-spin at one wing $B$ allows the experimenter at $B$ to
predict in advance the outcome of a quantum measurement of $z$-spin at the
distant wing $A$ (in ideal conditions). As was first argued by Einstein,
Podolsky and Rosen (1935) (using a somewhat different example), if locality is
assumed, then changing what is done at $B$ (from a $z$-spin measurement to
some other measurement) cannot affect the outcome at $A$, and therefore the
$z$-spin outcome at $A$ must be determined in advance regardless of what
measurement is performed at $B$. Having reached the conclusion that the
outcomes at $A$ and $B$ are locally determined, one can then run a Bell-type
argument, showing that their statistical correlation is incompatible with the
predicted (and observed) quantum correlation. If we leave aside the
many-worlds interpretation,\footnote{The Bell inequalities do not apply in the
many-worlds theory, because their derivation assumes that a quantum
measurement has only one outcome.} it follows that locality is incompatible
with quantum theory. Note that in this argument, determinism at each wing is
not assumed, but \textit{deduced} from the assumption of locality (Bell 1987: 143).

There is then strong evidence (again, if we leave aside the possibility of
many-worlds) that in the above set-up the physical processes at $A$ and $B$
are not independent, no matter how remote $A$ and $B$ may be from each other.
This raises the question of how such nonlocally-connected processes may be
embedded into the structure of standard relativistic spacetime.

It is sometimes suggested that, instead of accepting the existence of
superluminal influences, the whole issue could be avoided by assuming that our
classical spacetime is merely emergent. Now, it may well be true that
classical spacetime is emergent (for example from a deeper discrete
structure). However, this does not affect the issue at all. The EPR-Bell
correlations observed in the laboratory take place at macroscopic distances
(for example 12 m; Aspect \textit{et al}. 1982), involving photons with quite
ordinary energies (for example visible photons of wavelength $\sim500$ nm;
Aspect \textit{et al}. 1982). The detection events are recorded as taking
place in a region of space and time whose structure may be operationally
defined by macroscopic rods and clocks, in the laboratory where the experiment
is performed. There is no doubt that the structure of spacetime in that
laboratory, as defined by macroscopic rods and clocks, is to very high
accuracy well-described by standard relativity theory. One may then ask,
\textit{in the approximation} where the background spacetime is approximately
classical, how the events or outcomes recorded at $A$ and $B$ are to be
embedded in the background spacetime. Whatever the final theory underlying
spacetime (if there is one) turns out to be, this question must have an
answer, and the aim of this article is to provide one.

\subsection{Non-Equilibrium Hidden Variables}

We shall consider the issue from the standpoint of deterministic
hidden-variables theories. These provide a mapping from initial hidden
parameters $\lambda$ to final outcomes of quantum measurements. The mapping
depends on the macroscopic settings defining the experimental set-up. For
entangled quantum states, the mapping is nonlocal, in the sense that outcomes
at one wing depend on experimental settings at the distant wing (in at least
one direction; Bell 1964). Thus, the nonlocality is clearly present in the
underlying dynamics associated with the mapping. Instantaneous signalling is
not possible in such theories, however, \textit{provided} the initial hidden
parameters $\lambda$ have a special `quantum equilibrium' distribution
$\rho_{\mathrm{QT}}(\lambda)$. This distribution is chosen so that the
resulting statistics of quantum measurement outcomes agree with quantum theory.

As we shall discuss in section 3, once one is given a deterministic
hidden-variables theory for individual systems -- where mathematically the
theory is defined by the mapping from $\lambda$ to outcomes -- then there is
no conceptual reason why one should not consider the physics of more general
`non-equilibrium' distributions $\rho(\lambda)\neq\rho_{\mathrm{QT}}(\lambda
)$. For such distributions, nonlocality is present not only for individual
outcomes, but also at the statistical level: the marginal statistics at one
wing of an entangled state do (generically) depend on measurement settings at
the distant wing (in at least one direction). In such circumstances, with
$\rho(\lambda)\neq\rho_{\mathrm{QT}}(\lambda)$, practical nonlocal signalling
would be possible (Valentini 1991a,b, 1992, 1996, 2001, 2002a,b,c).

If one takes deterministic hidden-variables theories seriously, then one is
driven to conclude that our inability to send instantaneous signals is merely
an accident of our living in a time and place where the parameters $\lambda$
have the special distribution $\rho_{\mathrm{QT}}(\lambda)$, for which
statistical noise happens to erase (on average) the effects of nonlocality.
This state is roughly analogous to a state of global thermal equilibrium in
classical physics, in which it would be impossible to convert heat into work
(as this requires differences of temperature). In such a world -- in a state
of thermodynamic `heat death' -- the inability to convert heat into work is
not a law of physics, but rather a contingent feature of the state of thermal
equilibrium. Similarly, in our view, the absence of superluminal signalling in
our world is not a law of physics, but rather a contingent feature of the
state of quantum equilibrium (Valentini 1991a,b, 1992, 1996, 2001, 2002a,b,c).

Non-equilibrium deviations $\rho(\lambda)\neq\rho_{\mathrm{QT}}(\lambda) $
might have existed in the very early universe, with the relaxation
$\rho(\lambda)\rightarrow\rho_{\mathrm{QT}}(\lambda)$ taking place during the
violence of the big bang (Valentini 1991a,b, 1992, 1996, 2001, 2002a,b,c;
Valentini and Westman 2005). In effect, a hidden-variables analogue of
Boltzmann's `heat death' may have actually taken place in our observable
universe. However, relic cosmological particles that decoupled at sufficiently
early times might still be out of equilibrium today (Valentini 1996, 2001). It
has also been suggested that quantum non-equilibrium might be generated in
systems that are entangled with degrees of freedom located behind the event
horizon of a black hole (Valentini 2004a,b).

In any case it is certainly true that, from a hidden-variables perspective,
quantum theory is merely the phenomenological description of the statistics of
a special state with $\rho(\lambda)=\rho_{\mathrm{QT}}(\lambda)$. In
principle, there exists a wider (and explicitly nonlocal) physics of
non-equilibrium with $\rho(\lambda)\neq\rho_{\mathrm{QT}}(\lambda)$.

\section{Physical \textbf{Structure of Spacetime}}

The structure of spacetime at the most fundamental level should be defined in
terms of the physics at the most fundamental level. In a deterministic
hidden-variables theory, emergent properties of the quantum equilibrium state
(such as locality) have no fundamental status. The truly fundamental and
nonlocal physics is visible only in non-equilibrium. Therefore, a fundamental
appraisal of spacetime structure must be in terms of non-equilibrium physics,
taking into account instantaneous signalling.

\subsection{Kinematics and Dynamics}

This might seem problematic, it still being common among physicists to
describe superluminal effects as `acausal'. But superluminal signalling
violates causality -- that is, gives rise to backwards-in-time signals in some
frames -- \textit{if} one assumes a locally Minkowski structure for spacetime.
Historically, the Minkowski structure was developed for a local physics. If
Nature turns out to be nonlocal, then one should consider revising that structure.

This may seem an obvious point. Yet, many physicists tend to think of
Minkowski spacetime as a prior (`God-given') background or stage on which
physics takes place (at least locally, ignoring gravitation for the moment). A
common view is that laws such as Maxwell's equations possess Lorentz symmetry
`because' spacetime has a Minkowski structure. It is as if we were first given
the stage of spacetime, and afterwards we wrote laws on it. But one could
\textit{equally} take the view that spacetime has a Minkowski structure
`because' the known laws all have a Lorentz symmetry.\footnote{Arguably, these
are two different ways of saying the same thing. The kinematical structure of
spacetime cannot be disentangled from the dynamics taking place within it
(Brown 2005).} This would certainly be closer to the historical facts: first
one discovers certain symmetries in the behaviour of matter, then one
postulates a spacetime structure that incorporates those symmetries.

From this last perspective, one should be open to the possibility that, in the
future, new phenomena might break old symmetries, or, that new symmetries
might emerge; and in either case, the structure of spacetime might have to be
revised. In a word, one should bear in mind that new laws of physics might
demand a new structure for spacetime.

Kinematics and dynamics are two sides of the same coin (Brown 2005). As we
discover new dynamical effects, we should be prepared to modify our kinematics
(or spacetime geometry) if necessary or convenient. In section 3 we shall
describe an effect whose observation in the future is, in the author's
opinion, to be expected from a hidden-variables perspective, and which would,
we argue, lead us to modify our current relativistic kinematics.

The rise of relativity theory should have taught us the lesson that the
structure of spacetime is not \textit{a priori}, but depends on physics --
just as more recently, with the rise of quantum computing, we have come to
learn that the theory of computation is not \textit{a priori} but depends on
physics. Unfortunately, after 1905, the dogma of Newtonian spacetime was
quickly replaced by the dogma of (local) Minkowski spacetime.

The replacement of one rigid view by another was perhaps due in part to
Einstein's unfortunate `operational' presentation in his first relativity
paper of 1905 (Einstein 1905), which treated macroscopic rods and clocks as if
they were fundamental entities. This led to a widespread misunderstanding,
according to which the resulting kinematics was somehow logically inevitable,
when in fact it was highly contingent on properties of the physical dynamics
known at the time.

\subsection{Einstein and Poincar\'{e} in 1905}

Einstein himself acknowledged the conceptual mistake in his autobiographical
notes of 1949:

\begin{quotation}
`The theory [special relativity] .... introduces two kinds of physical things,
i.e., (1) measuring rods and clocks, (2) all other things, e.g., the
electro-magnetic field, the material point, etc. This, in a certain sense, is
inconsistent; strictly speaking measuring rods and clocks would have to be
represented as solutions of the basic equations (objects consisting of moving
atomic configurations), not, as it were, as theoretically self-sufficient
entities' (Einstein 1949b).
\end{quotation}

In 1905 Einstein had treated rods and clocks as primitive entities,
independent of theory (`theoretically self-sufficient'). But in fact, as
Einstein later recognised, rods and clocks are phenomenological entities
arising out of some underlying theory (perhaps involving particles and/or
fields). In reality, we need some body of theory to tell us how to construct
reliable rods and clocks and to analyse their behaviour. For example, using
theory we can calculate the effect of acceleration on a real clock, and so use
theory to design more robust clocks. Rods and clocks are not simply `given' to us.

The modern view of special relativity, used in high-energy physics for
example, makes no mention of rods and clocks. It concerns particles and fields
on Minkowski spacetime. The essence of Lorentz invariance is simply that the
Lagrangian density appearing in quantum field theory should be a Lorentz
scalar (resulting in a Lorentz-covariant $S$-matrix). Nor do classical light
waves play any special role: what matters are the symmetries of the
fundamental equations, not the speed of propagation of some particular
particle or field. After all, the photon might turn out to have a small mass.
That we first discovered Lorentz invariance via the classical electromagnetic
field is merely a historical accident, and Einstein's 1905 approach -- based
on macroscopic rods, clocks and classical light waves -- is merely a
historical (and fundamentally inconsistent) heuristic.

The popularity of Einstein's `operational' approach to special relativity had
the effect of introducing a deep and widespread confusion between
phenomenological and fundamental entities. This confusion seems to have
encouraged an overly-rigid philosophy of space and time, in which Einstein's
kinematics came to appear as an inevitable -- \textit{a priori}, and
theoretically self-sufficient -- background to the laws of
dynamics.\footnote{As we shall discuss elsewhere, the confusion between
phenomenology and fundamentals also led to inconsistencies in quantum theory,
in the form of the `measurement' or `reality' problem.} Today, despite the
discovery of quantum nonlocality, there is still a reluctance in some quarters
to even consider changing our view of spacetime structure.

It is often claimed that Einstein's 1905 approach should be regarded as not
merely a historical curiosity, but as the proper way to understand special
relativity. After all, it was this approach which in fact first led us to
special relativity. And how else could special relativity have been
discovered? But as a matter of historical fact, building on earlier work by
Lorentz and others, the formal structure of special relativity -- the
relativity principle, the universality of the Lorentz group, the relativistic
addition of velocities, and even 4-vectors with the associated 4-dimensional
invariant interval (later taken up by Minkowski in 1908) -- was independently
arrived at by Poincar\'{e} in his paper `On the Dynamics of the Electron'
(1906). This paper was submitted to a mathematical journal in Palermo, in the
same summer (of 1905) as Einstein's first relativity paper was submitted to
the \textit{Annalen der Physik}; it was published in 1906. A summary of the
results was published in 1905, in a short paper of the same title
(Poincar\'{e} 1905b).\footnote{According to its original title page,
Poincar\'{e}'s long paper `On the Dynamics of the Electron' (1906) was
accepted for publication by the \textit{Rendiconti del Circolo Matematico di
Palermo} on 23 July 1905, printed on 14 December, and officially published in
1906. The short summary with the same title was (according to Pais; 1982)
communicated to the Acad\'{e}mie des Sciences in Paris on 5 June 1905; it was
published in 1905 in the \textit{Comptes Rendus de l'Acad\'{e}mie des Sciences
de Paris }(Poincar\'{e} 1905b). Einstein's first relativity paper was received
by the \textit{Annalen der Physik} on 30 June 1905 and published in 1905
(Einstein 1905).}

The importance of Poincar\'{e}'s `Palermo' paper has been underestimated, even
by some historians. Certainly, most physicists are not even aware of its
existence. (An incomplete translation appears in Kilmister (1970); a
modernised presentation of most of the paper is given in Schwartz (1971,
1972). For detailed analyses of the paper, see Miller (1973) and Zahar (1989).
More recent discussions of Poincar\'{e}'s extensive contributions to special
relativity have been given by Darrigol (1995, 1996) and Granek (2000).)

Among physicists, Pauli was exceptional in being careful to credit
Poincar\'{e}'s Palermo paper properly throughout his celebrated treatise on
relativity (Pauli 1958; first published in 1921). For example, with reference
to the Palermo paper, Pauli notes that:

\begin{quotation}
`The formal gaps left by Lorentz's work were filled by Poincar\'{e}. He stated
the relativity principle to be generally and rigorously valid. Since he ....
assumed Maxwell's equations to hold for the vacuum, this amounted to the
requirement that all laws of nature must be covariant with respect to the
`Lorentz transformation'. The terms `Lorentz transformation' and `Lorentz
group' occurred for the first time in this paper by
Poincar\'{e}.\footnote{This sentence appears as a footnote in the original
text.} .... Poincar\'{e} further corrected Lorentz's formulae for the
transformations of charge density and current and so derived the complete
covariance of the field equations of electron theory' (Pauli 1958: 3).
\end{quotation}

Pauli correctly credits Poincar\'{e}, not only for postulating the Lorentz
group as a universal symmetry group, but also for the first use of 4-vectors
and of the associated 4-dimensional invariant interval. Pauli writes:

\begin{quotation}
`As a precursor of Minkowski one should mention Poincar\'{e} .... He already
introduced on occasion the imaginary coordinate $u=ict$ and combined, and
interpreted as point coordinates in $R_{4}$, those quantities which we now
call vector components. Furthermore, the invariant interval plays a r\^{o}le
in his considerations' (Pauli 1958: 21).
\end{quotation}

How had Poincar\'{e} done it? The answer is, along the lines that most workers
in high-energy physics would probably take today. (Poincar\'{e} was concerned
with the detailed structure and dynamics of the electron, the `elementary
particle physics' of the time.) He first notes the experimental failure to
detect the absolute motion of the Earth, and proposes that this is `a general
law of Nature', which he calls the `Relativity Postulate' (Poincar\'{e} 1906:
129). Further, following and perfecting the extensive work of Lorentz,
Poincar\'{e} notes that Maxwell's equations have the Lorentz group as an exact
symmetry group, and \textit{postulates} that this is a universal symmetry
applicable to all forces (including gravitation). Poincar\'{e} recognises that
this postulate suffices to explain the observed invariance of phenomena under
a boost. Citing Lorentz, Poincar\'{e} writes:

\begin{quotation}
`If one can impart a common boost to the whole system without any of the
apparent phenomena being modified, this is because the equations of an
electromagnetic medium are not changed by certain transformations, which we
shall call \textit{Lorentz transformations}; two systems, one at rest, the
other in motion, thus become exact images of each other. .... According to him
[Lorentz], all forces, whatever their origin, are affected by the Lorentz
transformation (and therefore by a boost) in the same manner as
electromagnetic forces' (Poincar\'{e} 1906: 130; translation by the author).
\end{quotation}

Poincar\'{e} then deduces the detailed structure of the Lorentz group,
including the relativistic addition of velocities, noting that the group
leaves invariant the quadratic form $x^{2}+y^{2}+z^{2}-t^{2}$. There follows
an extensive discussion of relativistic electron dynamics. In the final
section of the paper, Poincar\'{e} formulates a Lorentz-covariant
generalisation of Newtonian gravitation, with gravitational interactions
propagating at the speed of light\footnote{Poincar\'{e}'s short summary
(1905b) refers to `gravitational waves' propagating between gravitating
bodies. For a detailed discussion of Poincar\'{e}'s 1905 theory of gravity,
see Zahar (1989: 192--200).}. This last theory is formulated by finding
Lorentz-invariant functions of the velocities and relative positions of the
masses (as well as of time). To find these, Poincar\'{e} uses the fact that
the Lorentz group may be regarded as the group of rotations in a 4-dimensional
space with coordinates $x$, $y$, $z$, $it$. As Poincar\'{e} put it:

\begin{quotation}
`We see that the Lorentz transformation is nothing but a rotation of this
space around the origin' (Poincar\'{e} 1906: 168; translation by the author).
\end{quotation}

Independently of Einstein and Minkowski, then, in 1905 Poincar\'{e} arrived at
the formal, mathematical structure of Minkowski spacetime and the Lorentz group.

One may argue over the extent to which Poincar\'{e} understood the new
kinematics defined by his formalism. According to Darrigol (1995: 35, 1996:
280), Poincar\'{e} did understand that the Lorentz-transformed coordinates
were to be identified with the actual readings of boosted rods and clocks,
since he regarded Lorentz invariance as a physical (not just a mathematical)
symmetry, whereby `apparent phenomena' in a moving system follow the same laws
as phenomena in a system at rest. Similarly, according to Janssen and Stachel
(2004): `Unlike Lorentz, Poincar\'{e} realized that the auxiliary quantities
are the measured quantities for the moving observer'. In fact, as early as
1900, Poincar\'{e} understood that if experimenters moving with speed $v$ were
to assume that the speed of light is $c$ in every direction, then (to lowest
order in $v/c$) they would synchronise clocks separated by a distance $x$ such
that the settings differ by $-vx/c^{2}$ (see section 4.1). At least to lowest
order in $v/c$, Poincar\'{e} had already understood in 1900 that the
Lorentz-transformed time corresponded to the actual readings of moving
clocks.\footnote{Brown (2005), however, questions whether in 1905 Poincar\'{e}
fully understood the physical significance of the transformed coordinates to
higher orders in $v/c$. On the other hand, Darrigol (1995: 37--40) shows that,
in lectures delivered at the Sorbonne in 1906-07, Poincar\'{e} (apparently
independently of Einstein) generalised his 1900 discussion of clock
synchronisation (taking into account length contraction) to obtain the full
Lorentz-transformed time to all orders in $v/c$.}

Any suggestion that Poincar\'{e} viewed the Lorentz transformation as a purely
mathematical change of variables seems untenable. After all, Poincar\'{e}
asserted that Lorentz invariance alone sufficed to explain the invariance of
apparent phenomena under a boost, so the transformed quantities in question
must indeed have been regarded as those measured by a moving observer. (In
contrast, for Lorentz, his `theorem of corresponding states' -- which was
mathematically almost the same as Lorentz invariance -- had to be supplemented
by further physical assumptions to explain the failure to detect ether drift
(Janssen and Stachel 2004).) Further, in his Palermo paper, Poincar\'{e}
derives real physical corrections to Newton's law of gravity, from the
requirement that the law of motion for gravitating bodies should be covariant
with respect to rotations in what we would now call Minkowski space (with
coordinates $x$, $y$, $z$, $it$). For Poincar\'{e}, this symmetry clearly had
real, observable physical consequences.

One may also ask if Poincar\'{e} (like Lorentz) took the view that there was a
true rest frame. According to Darrigol (1995: 40), for example, Poincar\'{e}
did indeed maintain this view (which Darrigol sees as the only essential
difference between Poincar\'{e} and Einstein in 1905). On this point it should
be remembered that (as we shall discuss in section 5) for Poincar\'{e}, the
geometry of spacetime is not a fact about the world but merely a convenient
convention, so that if one finds it convenient one may indeed think in terms
of absolute space and time.\footnote{We are inclined to agree with Zahar
(1989: 150): `.... that Poincar\'{e} did discover special relativity, that his
philosophy of science provided him with heuristic guidelines, but that certain
ambiguities within that same philosophy prevented both his contemporaries and
many historians from appreciating the true value of his contribution.'} In any
case, this interpretation of Poincar\'{e}'s made no empirical difference.
Further, we argue that in the light of quantum nonlocality it may well be the
better interpretation after all.

It does seem fair to say -- despite (limited) anticipations by Fitzgerald,
Lorentz, and Larmor\footnote{A limited form of time dilation was anticipated
by Larmor (in a paper of 1897, and in his book of 1900), and by Lorentz (in a
paper of 1899). See Brown (2005: section 4.5), and Janssen and Stachel
(2004).} -- that a clear and complete statement of universal time dilation and
length contraction is first found in Einstein's paper of 1905. Poincar\'{e}'s
Palermo paper discusses length contraction for spherical electrons, but does
not explicitly mention time dilation, despite extensive use of the
Lorentz-transformed time variable. As Pauli observed, regarding time dilation:

\begin{quotation}
`While this consequence of the Lorentz transformation was already implicitly
contained in Lorentz's and Poincar\'{e}'s results, it received its first clear
statement only by Einstein' (Pauli 1958: 13).
\end{quotation}

It was claimed by Pais (1982: 164, 167--168) that even after 1905 Poincar\'{e}
did not understand special relativity, because, judging from the text of his
lectures at G\"{o}ttingen in 1909 (Poincar\'{e} 1910), he did not understand
that length contraction was a consequence of Einstein's two postulates (the
relativity principle and the light postulate), but instead insisted on
including length contraction as a third postulate. In the author's opinion,
this issue is confused because Einstein's 1905 approach actually contains an
implicit third postulate: that under a boost from one rest frame to another,
unit rods are transformed into unit rods, and similarly for unit clock ticks.
Einstein himself admitted this, in a footnote to a review he published in
1910, where he writes:

\begin{quotation}
`It should be noted that we will always implicitly assume that the fact of a
measuring rod or a clock being set in motion or brought to rest does not
change the length of the rod or the rate of the clock' (Einstein 1993: 130).
\end{quotation}

To the author's knowledge, the only other place in the historical literature
where Einstein's implicit third postulate is mentioned is in Born's relativity
text (1962).\footnote{The recent book by Brown (2005: section 2.4) calls this
assumption the `boostability' of rods and clocks, and regards it more as a
`stipulation' (or convenient convention) than an assumption.} In fact, Born
discusses this postulate in some detail, and regards it as of crucial
importance. He writes:

\begin{quotation}
`.... it is assumed as self-evident that a measuring rod which is brought into
one system of reference $S$ and then into another $S^{\prime}$ under exactly
the same physical conditions would represent the same length in each .... A
fixed rod that is at rest in the system $S$ and is of length 1 cm. will, of
course, also have the length 1 cm. when it is at rest in the system
$S^{\prime}$ .... Exactly the same would be postulated for the clocks .... We
might call this tacit assumption of Einstein's theory the \textquotedblleft
principle of the physical identity of the units of measure\textquotedblright%
\ .... This is the feature of Einstein's theory by which it rises above the
standpoint of a mere convention and asserts definite properties of real
bodies' (Born 1962: 251--252).
\end{quotation}

It might be thought that the third postulate could be dispensed with, by using
the relativity principle to deduce that any specific process for constructing
rods and clocks must give the same results in all inertial frames. Certainly,
using the light postulate as well, one could then deduce that the Lorentz
transformation relates the readings of \textit{different} rods and clocks that
have been constructed (by a similar process) in different inertial frames.
However, one would still have deduced nothing about what happens when the
\textit{same} rod or clock is boosted (or accelerated) from one inertial frame
to another. (As an example one might, in principle, envisage a theory
satisfying the relativity principle and the light postulate, but with the
additional property that once a rod or clock has been constructed in a given
inertial frame it is destroyed by any subsequent arbitrarily small acceleration.)

Thus, despite widespread opinion to the contrary, length contraction and time
dilation under a boost do \textit{not} follow from Einstein's two postulates
alone. A further postulate is required, to relate the readings of rods and
clocks boosted from one inertial frame to another.\footnote{In fact, a still
further assumption of spatial isotropy is also needed -- see Brown (2005:
section 5.4.3).}

In view of the crucial importance of the third assumption implicitly used by
Einstein, it must be regarded as regrettable that Einstein did not mention it
explicitly in his first relativity paper. In the author's opinion, it is quite
possible that Poincar\'{e} was aware of this lacuna, explaining why in his
lectures of 1909 (Poincar\'{e} 1910) -- where he sketches an axiomatic basis
for the `new mechanics', in terms of simple physical postulates independent of
the details of Maxwell's equations, much as Einstein did in 1905 -- he added
the third postulate of length contraction, which was not as elegant as the
third postulate implicitly used by Einstein, but effective
nonetheless.\footnote{As noted by Darrigol (1995: 39), Poincar\'{e} had also
used length contraction as a hypothesis in his Sorbonne lectures of 1906-07.
Again, in the author's view, Poincar\'{e} may well have understood that some
such extra hypothesis was needed to relate measurements in different frames.}

In any case, such detailed questions of priority, or of who understood exactly
what and when, while historically interesting, are not strictly relevant here.
What really matters, for our purpose, is that the approach taken in
Poincar\'{e}'s Palermo paper -- in which the Lorentz group is first discovered
through Maxwell's equations and then postulated to be a universal (physical)
symmetry group -- quite plainly \textit{could have been }the historical route
to special relativity. Regardless of the extent to which Poincar\'{e} did or
did not understand it at the time, the fact is that the kinematics of
Minkowski spacetime was contained in the formal structure put forward in
Poincar\'{e}'s paper.

Minkowski, in his famous lecture on `Space and Time' delivered in 1908
(Minkowski 1952), appears to express a preference for this sort of approach,
which actually goes back to 1887 when Voigt (1887) derived the Lorentz
transformation, up to an overall constant factor, as a symmetry of the
(scalar) wave equation.\footnote{Voigt's paper is briefly discussed by Pais
(1982: 121--122), and in great detail by Ernst and Hsu (2001), who also
provide an English translation of it.} According to Minkowski:

\begin{quotation}
`Now the impulse and true motive for assuming the group $\mathrm{G}_{c}$ [that
is, the Poincar\'{e} group, which leaves invariant the 4-dimensional interval]
came from the fact that the differential equation for the propagation of light
in empty space possesses that group $\mathrm{G}_{c}$. An application of this
fact in its essentials has already been given by W. Voigt, G\"{o}ttinger
Nachrichten, 1887, p. 41\footnote{This sentence appears as a footnote in the
original text.}' (Minkowski 1952: 81).
\end{quotation}

It is sometimes argued that Einstein's operational approach has the advantage
of being independent of the details of specific equations such as Maxwell's.
This may be so, but Einstein's approach also has the disadvantage of giving a
special status to classical light waves, and of being conceptually
inconsistent with regard to the nature of rods and clocks. As for
Poincar\'{e}'s approach, as a scientific methodology there is nothing wrong
with discovering a symmetry in certain equations and then postulating that the
symmetry is universal (regardless of whether those equations turn out to be
fundamental or not). This is, after all, common practice in high-energy
physics today. Clearly, Einstein's operational approach was not necessary, and
special relativity could have been (and arguably essentially was) discovered
without appeal to a fundamentally inconsistent operationalism.

In the author's opinion, if Poincar\'{e}'s approach had in fact been the
generally-accepted historical route to special relativity, then physicists
today might be more keenly aware that spacetime geometry is not `prior to'
dynamics but rather a reflection of symmetries of the currently-known
dynamics.\footnote{Even if Poincar\'{e} himself, for philosophical reasons of
his own, seemed to prefer retaining the old notions of space and time in the
background.} From this standpoint, as physics progresses, the structure of
spacetime is as subject to possible revision as are the laws of dynamics themselves.

\section{Instantaneous Signalling in Quantum Non-Equilibrium}

In this section we show how, in deterministic hidden-variables theories, a
non-standard distribution of hidden variables (generically) gives rise to
instantaneous signalling at the statistical level. We first discuss this for
general theories (Valentini 2002a,b), then for the specific example of the
pilot-wave theory of de Broglie and Bohm (Valentini 1991b, 2002c).

\subsection{General (Deterministic) Hidden-Variables Theories}

For a 2-state system, consider quantum observables of the form $\hat{\sigma
}=\mathbf{m}\cdot\mathbf{\hat{\sigma}}$, where $\mathbf{m}$ is a unit vector
specifying a point on the Bloch sphere and $\mathbf{\hat{\sigma}}$ is the
Pauli spin operator. The values $\sigma=\pm1$ are obtained upon performing a
quantum measurement of $\hat{\sigma}$. Over an ensemble with density operator
$\hat{\rho}$, the quantum expectation value of $\hat{\sigma}$ is given by the
Born rule as $\left\langle \hat{\sigma}\right\rangle =\mathrm{Tr}\left(
\hat{\rho}\mathbf{m}\cdot\mathbf{\hat{\sigma}}\right)  =\mathbf{m}%
\cdot\mathbf{P}$, where $\mathbf{P}=\langle\mathbf{\hat{\sigma}}\rangle$ (with
norm $0\leq P\leq1$) is the mean polarisation. The quantum probabilities
$p_{\mathrm{QT}}^{\pm}(\mathbf{m})$ for outcomes $\sigma=\pm1$ are then fixed
as%
\begin{equation}
p_{\mathrm{QT}}^{\pm}(\mathbf{m})=\frac{1}{2}\left(  1\pm\mathbf{m}%
\cdot\mathbf{P}\right)  \label{Eqp}%
\end{equation}

In a (deterministic) hidden-variables theory, for every run of the experiment
with measurement axis $\mathbf{m}$, there are hidden parameters collectively
denoted $\lambda$ that determine the outcome $\sigma=\pm1$ according to some
mapping $\sigma=\sigma\left(  \mathbf{m},\lambda\right)  $. Over an ensemble
of experiments, the observed distribution of outcomes is explained by some
assumed distribution $\rho_{\mathrm{QT}}(\lambda)$ of parameters $\lambda$,
where $\rho_{\mathrm{QT}}(\lambda)$ is such that expectations%
\[
\left\langle \sigma\left(  \mathbf{m},\lambda\right)  \right\rangle
_{\mathrm{QT}}=\int d\lambda\ \rho_{\mathrm{QT}}(\lambda)\sigma\left(
\mathbf{m},\lambda\right)
\]
agree with the quantum prediction $\left\langle \mathbf{m}\cdot\mathbf{\hat
{\sigma}}\right\rangle $. The values of $\lambda$ are usually defined at some
initial time, say at the time of preparation of the quantum state. The
outcomes $\sigma=\sigma\left(  \mathbf{m},\lambda\right)  $ are defined at the
time of measurement.

Now, there is a clear conceptual distinction between the initial values
$\lambda$ and the mapping $\sigma=\sigma\left(  \mathbf{m},\lambda\right)  $
to final outcomes $\sigma$. In particular, the former amount to what are
usually called `initial conditions', while the latter would usually be called
a `dynamical law' that maps initial conditions to final states. Therefore,
once such a theory has been constructed, one may contemplate arbitrary initial
conditions -- over an ensemble, distributions $\rho(\lambda)\neq
\rho_{\mathrm{QT}}(\lambda)$ -- while retaining the mapping $\sigma
=\sigma\left(  \mathbf{m},\lambda\right)  $. Generically, such `non-quantum'
or `non-equilibrium' distributions will yield expectation values%
\[
\left\langle \sigma\left(  \mathbf{m},\lambda\right)  \right\rangle =\int
d\lambda\ \rho(\lambda)\sigma\left(  \mathbf{m},\lambda\right)
\]
that disagree with quantum theory, and the statistics of outcomes will
generally violate the standard quantum-theoretical constraints. Note the key
conceptual point: we have the same deterministic mapping $\sigma=\sigma\left(
\mathbf{m},\lambda\right)  $ for each system, regardless of the (equilibrium
or non-equilibrium) distribution for the ensemble.

Many of the supposedly fundamental constraints of quantum theory, such as
statistical locality, are (from a hidden-variables perspective) merely
contingent features of the special distribution $\rho_{\mathrm{QT}}(\lambda)$.
As noted in section 1.3, there is an analogy here with the contingent
constraints that arise in classical physics in a state of global thermal
equilibrium: the inability to convert heat into work is not fundamental, but a
contingency due to all systems having the same temperature.

Consider a pair of widely-separated 2-state systems with spatial locations $A
$ and $B$. Quantum measurements of $\hat{\sigma}_{A}\equiv\mathbf{m}_{A}%
\cdot\mathbf{\hat{\sigma}}_{A}$, $\hat{\sigma}_{B}\equiv\mathbf{m}_{B}%
\cdot\mathbf{\hat{\sigma}}_{B}$ can yield outcomes $\sigma_{A}$, $\sigma
_{B}=\pm1$. For the singlet state%
\[
\left\vert \Psi\right\rangle =\left(  \left\vert +\mathbf{n},-\mathbf{n}%
\right\rangle -\left\vert -\mathbf{n},+\mathbf{n}\right\rangle \right)
/\surd2
\]
(for any axis $\mathbf{n}$) quantum theory predicts that outcomes $\sigma_{A}%
$, $\sigma_{B}=\pm1$ occur in the ratio $1:1$ at each wing, with a correlation%
\begin{equation}
\left\langle \Psi\right\vert \hat{\sigma}_{A}\hat{\sigma}_{B}\left\vert
\Psi\right\rangle =-\mathbf{m}_{A}\cdot\mathbf{m}_{B} \label{correln}%
\end{equation}
Nevertheless, the distant settings have no effect on the expectation values
($\left\langle \hat{\sigma}_{A,B}\right\rangle =0$) or on the probabilities
($p_{\mathrm{QT}}^{\pm}(\mathbf{m}_{A,B})=1/2$) at each wing, making nonlocal
signalling impossible.

However, from a hidden-variables perspective, Bell's theorem (1964) tells us
that to reproduce this correlation a hidden-variables theory must take the
nonlocal form%
\begin{equation}
\sigma_{A}=\sigma_{A}(\mathbf{m}_{A},\mathbf{m}_{B},\lambda),\;\;\;\;\sigma
_{B}=\sigma_{B}(\mathbf{m}_{A},\mathbf{m}_{B},\lambda) \label{deteqns}%
\end{equation}
in which the individual outcomes $\sigma_{A}$, $\sigma_{B}$ do depend on the
distant measurement settings. Only with such nonlocal dependence can the
theory reproduce the quantum correlation%
\begin{equation}
\left\langle \sigma_{A}\sigma_{B}\right\rangle _{\mathrm{QT}}\equiv\int
d\lambda\ \rho_{\mathrm{QT}}(\lambda)\sigma_{A}(\mathbf{m}_{A},\mathbf{m}%
_{B},\lambda)\sigma_{B}(\mathbf{m}_{A},\mathbf{m}_{B},\lambda)=-\mathbf{m}%
_{A}\cdot\mathbf{m}_{B} \label{qucorreln}%
\end{equation}
for some ensemble distribution $\rho_{\mathrm{QT}}(\lambda)$. More precisely,
at least one of $\sigma_{A}$, $\sigma_{B}$ must depend on the distant setting,
and without loss of generality we shall assume that $\sigma_{A}$ has a
nonlocal dependence on $\mathbf{m}_{B}$.

Now, for an arbitrary ensemble with $\rho(\lambda)\neq\rho_{\mathrm{QT}%
}(\lambda)$, in general%
\begin{equation}
\left\langle \sigma_{A}\sigma_{B}\right\rangle \equiv\int d\lambda
\ \rho(\lambda)\sigma_{A}(\mathbf{m}_{A},\mathbf{m}_{B},\lambda)\sigma
_{B}(\mathbf{m}_{A},\mathbf{m}_{B},\lambda)\neq-\mathbf{m}_{A}\cdot
\mathbf{m}_{B} \label{nonqucorreln}%
\end{equation}
and the outcomes $\sigma_{A}$, $\sigma_{B}=\pm1$ at each wing will occur in a
ratio generally differing from $1:1$. Further, under a change in the
measurement setting at one wing, the outcome statistics at the distant wing
will generally change, amounting to a nonlocal signal at the statistical
level. The key point here is that, assuming a nonlocal dependence of
$\sigma_{A}$ on $\mathbf{m}_{B}$, the `transition sets'%
\begin{align*}
T_{A}(-,+)  &  \equiv\left\{  \lambda|\sigma_{A}(\mathbf{m}_{A},\mathbf{m}%
_{B},\lambda)=-1,\;\sigma_{A}(\mathbf{m}_{A},\mathbf{m}_{B}^{\prime}%
,\lambda)=+1\right\} \\
T_{A}(+,-)  &  \equiv\left\{  \lambda|\sigma_{A}(\mathbf{m}_{A},\mathbf{m}%
_{B},\lambda)=+1,\;\sigma_{A}(\mathbf{m}_{A},\mathbf{m}_{B}^{\prime}%
,\lambda)=-1\right\}
\end{align*}
cannot be empty for arbitrary settings $\mathbf{m}_{A}$, $\mathbf{m}_{B}$,
$\mathbf{m}_{B}^{\prime}$. Some outcomes at $A$ must change under a shift
$\mathbf{m}_{B}\rightarrow\mathbf{m}_{B}^{\prime}$ at $B$. In quantum
equilibrium, the ratio of outcomes $\sigma_{A}=\pm1$ is $1:1$ for all
settings, therefore we must have `detailed balancing'%
\[
\mu_{\mathrm{QT}}[T_{A}(-,+)]=\mu_{\mathrm{QT}}[T_{A}(+,-)]
\]
with respect to the equilibrium measure $d\mu_{\mathrm{QT}}\equiv
\rho_{\mathrm{QT}}(\lambda)d\lambda$. In other words, in quantum equilibrium,
the fraction of the ensemble making the transition $\sigma_{A}=-1\rightarrow
\sigma_{A}=+1$ under $\mathbf{m}_{B}\rightarrow\mathbf{m}_{B}^{\prime}$ must
equal the fraction making the reverse transition $\sigma_{A}=+1\rightarrow
\sigma_{A}=-1$. (This is analogous to the principle of detailed balance in
statistical mechanics: thermal equilibrium is maintained if the mean
transition rate from state $i$ to state $j$ is equal to the mean transition
rate from $j$ to $i$.) Since $T_{A}(-,+)$ and $T_{A}(+,-)$ are fixed by
deterministic equations, they are independent of the ensemble distribution of
$\lambda$. Thus, for a hypothetical non-equilibrium ensemble $\rho
(\lambda)\neq\rho_{\mathrm{QT}}(\lambda)$, in general%
\[
\mu\lbrack T_{A}(-,+)]\neq\mu\lbrack T_{A}(+,-)]
\]
where $d\mu\equiv\rho(\lambda)d\lambda$. In other words, the fraction of the
non-equilibrium ensemble making the transition $\sigma_{A}=-1\rightarrow
\sigma_{A}=+1$ will not in general balance the fraction making the reverse
transition; the ratio of outcomes at $A$ will in general change under
$\mathbf{m}_{B}\rightarrow\mathbf{m}_{B}^{\prime}$ and there will be
instantaneous signals at the statistical level from $B$ to $A$ (Valentini 2002a,b).

In any deterministic hidden-variables theory, then, hypothetical
non-equilibrium distributions $\rho(\lambda)\neq\rho_{\mathrm{QT}}(\lambda)$
generally make it possible to use nonlocality for instantaneous signalling
(just as, in ordinary statistical physics, differences of temperature make it
possible to convert heat into work) (Valentini 2002a,b).

\subsection{The Example of Pilot-Wave Theory}

Non-equilibrium signalling at a distance was first noted (Valentini 1991b,
2002c) in the hidden-variables theory of de Broglie and Bohm (de Broglie 1928;
Bohm 1952a,b). In this `pilot-wave theory' (as it was originally called by de
Broglie), a system with wave function $\Psi(X,t)$ satisfying the
Schr\"{o}dinger equation%
\begin{equation}
i\frac{\partial\Psi}{\partial t}=\hat{H}\Psi\label{Sch0}%
\end{equation}
has an actual configuration $X(t)$ whose motion is given by the first-order
differential equation%
\begin{equation}
\dot{X}(t)=\frac{J(X,t)}{\left\vert \Psi(X,t)\right\vert ^{2}} \label{deB0}%
\end{equation}
where $J=J\left[  \Psi\right]  =J(X,t)$ (which in quantum theory is called the
`probability current') satisfies the continuity equation%
\begin{equation}
\frac{\partial\left\vert \Psi\right\vert ^{2}}{\partial t}+\nabla_{X}\cdot J=0
\label{Cont0}%
\end{equation}
(which follows from (\ref{Sch0})). In pilot-wave theory, $\Psi$ is regarded as
an objective physical field guiding the system.

For example, for a system of $N$ particles with 3-vector positions
$\mathbf{x}_{i}(t)$ and masses $m_{i}$ ($i=1,2,....,N)$ the wave function
$\Psi(X,t)$ on $3N$-dimensional configuration space ($X\equiv(\mathbf{x}%
_{1},\mathbf{x}_{2},....,\mathbf{x}_{N})$) is a complex field obeying the
Schr\"{o}dinger equation%
\begin{equation}
i\frac{\partial\Psi}{\partial t}=\sum_{i=1}^{N}-\frac{1}{2m_{i}}\nabla_{i}%
^{2}\Psi+V\Psi\label{Sch1}%
\end{equation}
and the particle velocities are given by%
\begin{equation}
\frac{d\mathbf{x}_{i}}{dt}=\frac{1}{m_{i}}\operatorname*{Im}\left(
\frac{\mathbf{\nabla}_{i}\Psi}{\Psi}\right)  =\frac{\mathbf{\nabla}_{i}%
S}{m_{i}} \label{deB1}%
\end{equation}
where $\Psi=\left\vert \Psi\right\vert e^{iS}$ and we take $\hbar=1$.

Equations (\ref{Sch0}) and (\ref{deB0}) determine the motion $X(t)$ of an
\textit{individual} system, given the initial configuration $X(0)$ and wave
function $\Psi(X,0)$ at $t=0$. If we are given an arbitrary initial
distribution $P(X,0)$, for an ensemble of systems with the same wavefunction
$\Psi(X,0)$, then the evolution of $P(X,t)$ is necessarily given by the
continuity equation%
\begin{equation}
\frac{\partial P}{\partial t}+\nabla_{X}\cdot(P\dot{X})=0 \label{Cont1}%
\end{equation}
This same equation is satisfied by $\left\vert \Psi\right\vert ^{2}$, as
follows from (\ref{Cont0}). Thus, if $P(X,0)=\left\vert \Psi(X,0)\right\vert
^{2}$ at some initial time, then $P(X,t)=\left\vert \Psi(X,t)\right\vert ^{2}$
at all times $t$. As shown by Bohm (1952a,b), one then recovers the
statistical predictions of quantum theory.

In pilot-wave theory, the outcome obtained in a given experiment is determined
by $X(0)$ and $\Psi(X,0)$, so that one may identify $\lambda$ with the pair
$X(0)$, $\Psi(X,0)$. For an ensemble of experiments with the same $\Psi(X,0)$,
in effect $\lambda$ is just $X(0)$, and the distribution $\rho_{\mathrm{QT}%
}(\lambda)$ is given by $P_{\mathrm{QT}}(X,t)=\left\vert \Psi(X,t)\right\vert
^{2}$. As in the general discussion above, we may retain the same
deterministic dynamics for individual systems, and consider a non-standard
distribution of initial conditions. Here, this means we retain the dynamical
equations (\ref{Sch0}), (\ref{deB0}) and consider an arbitrary initial
ensemble with $P(X,0)\neq\left\vert \Psi(X,0)\right\vert ^{2}$. The evolution
of $P(X,t)$ will be determined by (\ref{Cont1}).

In appropriate circumstances, (\ref{Cont1}) leads to relaxation $P\rightarrow
\left\vert \Psi\right\vert ^{2}$ on a coarse-grained level (Valentini 1991a,
1992, 2001; Valentini and Westman 2005), much as the corresponding classical
evolution on phase space leads to thermal relaxation. However, for as long as
the ensemble is in non-equilibrium, the statistics of outcomes of quantum
measurements will disagree with quantum theory.

As required by Bell's theorem, pilot-wave theory is fundamentally nonlocal.
For two particles whose wave function $\Psi(\mathbf{x}_{A},\mathbf{x}_{B},t) $
is entangled, $\mathbf{\dot{x}}_{A}(t)=\mathbf{\nabla}_{A}S(\mathbf{x}%
_{A},\mathbf{x}_{B},t)/m_{A}$ depends instantaneously on $\mathbf{x}_{B}$, and
ordinary operations on particle $B$ -- such as switching on a local potential
-- have an instantaneous effect on the motion of particle $A$. But for a
quantum equilibrium ensemble $P(\mathbf{x}_{A},\mathbf{x}_{B},t)=|\Psi
(\mathbf{x}_{A},\mathbf{x}_{B},t)|^{2}$, such operations on particle $B$ have
no statistical effect on particle $A$: the individual nonlocal effects are
masked by quantum noise.

As in the general case discussed above, nonlocality is (generally speaking)
hidden by statistical noise only in quantum equilibrium. For an ensemble of
entangled particles with initial distribution $P(\mathbf{x}_{A},\mathbf{x}%
_{B},0)\neq|\Psi(\mathbf{x}_{A},\mathbf{x}_{B},0)|^{2}$, a local change in the
Hamiltonian of particle $B$ generally induces an instantaneous change in the
marginal distribution $p_{A}(\mathbf{x}_{A},t)\equiv%
{\textstyle\int}
d^{3}\mathbf{x}_{B}\ P(\mathbf{x}_{A},\mathbf{x}_{B},t)$ of particle $A$. For
example, in one dimension, a sudden change $\hat{H}_{B}\rightarrow\hat{H}%
_{B}^{\prime}$ in the Hamiltonian of particle $B$ induces a change $\Delta
p_{A}\equiv p_{A}(x_{A},t)-p_{A}(x_{A},0)$ of the form (for small $t$)
(Valentini 1991b)%
\begin{equation}
\Delta p_{A}=-\frac{t^{2}}{4m}\frac{\partial}{\partial x_{A}}\left(
a(x_{A})\int dx_{B}\ b(x_{B})\frac{P(x_{A},x_{B},0)-|\Psi(x_{A},x_{B},0)|^{2}%
}{|\Psi(x_{A},x_{B},0)|^{2}}\right)  \label{Superc}%
\end{equation}
(Here $m_{A}=m_{B}=m$, the factor $a(x_{A})$ depends on $\Psi(x_{A},x_{B},0)
$, while $b(x_{B})$ also depends on $\hat{H}_{B}^{\prime}$ and vanishes if
$\hat{H}_{B}^{\prime}=\hat{H}_{B}$.) In general, the signal is non-zero if
$P_{0}\neq|\Psi_{0}|^{2}$ (that is, if $\rho(\lambda)\neq\rho_{\mathrm{QT}%
}(\lambda)$).

Elsewhere (Valentini 2002c), using the example of pilot-wave theory, we have
described how non-equilibrium particles might be detected in practice, by the
statistical analysis of random samples (taken, for example, from a parent
population of relic particles left over from the early universe). Once such
particles have been identified, they may be used as a resource for
superluminal signalling; further, they may be used to perform `subquantum
measurements' on ordinary, equilibrium systems (Valentini 2002c).

\section{Absolute Simultaneity in Flat and Curved Spacetime}

We have seen that, even at ordinary laboratory distances and energies, quantum
non-equilibrium would unleash instantaneous signals between entangled systems.
This raises the question of how these signals could mesh with the surrounding
approximately classical spacetime. As we emphasised in the Introduction, this
question must have an answer, \textit{irrespective} of the underlying
microscopic theory of spacetime.

If experimenters at spacetime events $A$ and $B$ had access to non-equilibrium
systems entangled between $A$ and $B$, then they would be able to signal back
and forth to each other instantaneously. In an arbitrarily short time (as
measured at each wing), a long conversation could in principle take place,
during which (for example) the experimenters agree to set their clocks to read
time $t=0$. They could signal to each other to confirm that they have done so.
In such conditions, $A$ and $B$ must be regarded as simultaneous events, and
the agreed-upon time variable $t$ would define an absolute simultaneity. Thus,
using non-equilibrium matter, experimenters at remote locations could set
their clocks to read the same instantaneous time $t$.

There are, however, some differences depending on whether gravitation is
absent or present. Let us discuss these in turn.

\subsection{Flat Spacetime}

In the absence of gravitation (where the kinematics is usually represented by
flat Minkowski spacetime), remote experimenters may use entangled
non-equilibrium systems to set their clocks to read the same time $t$.
However, they must be careful to bear in mind that clocks in motion drift out
of synchronisation with clocks at rest. For if a clock undergoes a spatial
displacement $d\mathbf{x}$ in a time $dt$, then the `proper' time $d\tau$
ticked by the clock is given by%
\[
d\tau^{2}=dt^{2}-d\mathbf{x}^{2}%
\]
Thus, a clock moving through space with speed $v=\left\vert d\mathbf{x}%
/dt\right\vert $ is slowed by the factor $1/\sqrt{1-v^{2}/c^{2}}$. Here, this
`time dilation' may be regarded as a dynamical effect of motion on the rate of
evolution of physical systems, as originally anticipated by Larmor and
Lorentz. (An instructive account of this viewpoint was given by Bell (1987: 67--80).)

One must then distinguish between \textit{simultaneity} and
\textit{synchronicity}. The first refers to events that exist `in unison', in
a sense that could be verified by nonlocal communication. The second refers
merely to the coincidence of readings of certain (usually classical,
macroscopic) systems called `clocks', where for dynamical reasons the rate of
evolution of such systems depends on how fast they are moving through space.
Simultaneity is not equivalent to synchronicity.

For example, if two clocks are initially close together and synchronised in a
standard inertial frame with time function $t$, and if one clock is
accelerated and eventually returns close to its partner, then finally the two
clocks will be out of step, as the accelerated clock will have been slowed
down. If $t$ coincides with our absolute time, the final clock readings will
correspond to simultaneous events, yet, the readings will not be synchronous.

Note that this dynamical effect of motion occurs at the classical macroscopic
level, as well as at the statistical level for ensembles of microscopic
quantum systems, but it is not necessarily relevant to the deeper level of
hidden variables. (For example, decay rates for individual atoms are affected
by time dilation, but such rates apply to quantum ensemble averages and not to
individual systems.) Therefore, there is no reason why this dynamical effect
should be built into the fundamental kinematics (as it usually is).

The objection might be raised that superluminal signals in a given frame would
`violate causality', since in other frames the signals could travel backwards
in time, leading to paradoxes. But as we discussed in section 2, this argument
assumes that the structure of spacetime is fundamentally Minkowskian. There is
no reason to assume this. At the nonlocal hidden-variable level, there may
well be a preferred slicing of spacetime, with a time function $t$ that
defines a fundamental causal sequence (Popper 1982; Bohm and Hiley 1984; Bell
1986, 1987).

Clearly, in a given preferred frame with standard Lorentzian coordinates $t$,
$x$, $y$, $z$, instantaneous signalling between distant experimenters would
not in itself be problematic. But what about the Lorentz transformation? One
might be disturbed by the idea that an experimenter moving along (for example)
the $x$-axis could `see' such signals propagating `backwards in time'.
However, a real experimenter does not simply `see' the global time of his
Lorentz frame. Rather, the experimenter has a collection of clocks distributed
over space, which have to be \textit{set} according to some chosen procedure.
The time associated with an event occurring at some point in space is just the
reading of the clock in the neighbourhood of that event. If an event $B$ is
for some physical reason regarded as `causing' a spatially-distant event $A$
(for example a message is sent from $B$ to $A$), and if the reading of a clock
at $B$ is larger than the reading of a clock at $A$, then before declaring
this paradoxical one ought to ask how the clocks at $A$ and $B$ were set in
the first place.

If the moving experimenter chooses Einstein's so-called `synchronisation',
using light pulses whose speed is taken to be isotropic, then at (preferred)
time $t$ the moving clock located at $x$, $y$, $z$ will read a time%
\begin{equation}
t%
\acute{}%
=\frac{t-vx/c^{2}}{\sqrt{1-v^{2}/c^{2}}} \label{tprime}%
\end{equation}
From our perspective, the interpretation of this formula is very simple. The
moving clocks distributed along $x>0$ have been initially set (for example at
$t=0$) to read progressively earlier times, with a lag proportional to $x$;
while the moving clocks along $x<0$ have been similarly set to read later
times. These settings have been chosen precisely so as to make a light pulse
(with speed $c$ in the preferred frame\footnote{The speed $c$ in the preferred
frame will of course be independent of the motion of the source, as expected
of a wave phenomenon.}) \textit{appear} to have a speed $c$, along both $+x$
and $-x$, in the moving frame. This is the origin of the term $-vx/c^{2}$ (to
lowest order in $v/c$). If one includes the effect of motion, which as we have
said slows clocks down, one also obtains the factor $1/\sqrt{1-v^{2}/c^{2}}$.

If the moving experimenter adopts the Einstein convention for the
synchronisation of clocks, then the settings (\ref{tprime}) have the following
peculiarity: an instantaneous signal propagating along $+x$ in the preferred
frame appears to be going `backwards in time' as judged by the moving clocks
with settings $t%
\acute{}%
$. That is, if the signal starts at $x_{B}$ and propagates to $x_{A}>x_{B}$,
then if $v>0$ the readings $t%
\acute{}%
_{A}$, $t%
\acute{}%
_{B}$ of the moving clocks at the events $A$, $B$ have the property $t%
\acute{}%
_{B}>t%
\acute{}%
_{A}$. But there is nothing mysterious or paradoxical here: for the moving
clocks were initially set with a time-lag proportional to $x$, and the result
$t%
\acute{}%
_{B}>t%
\acute{}%
_{A}$ is a direct and immediate reflection of this initial set-up. Indeed,
this phenomenon is exactly the same as the familiar `jet lag' which occurs
when an experimenter moves rapidly from one time zone to another on the
Earth's surface. Clocks distributed over the Earth's surface have been set
according to a convention related to the locally-observed position of the Sun
in the sky, and it is in no way surprising or problematic that a jet passenger
may in a formal sense `travel backwards in time'.

Note that, from this point of view, time dilation is a real physical effect of
motion which may be unambiguously verified by experiment (for example by
taking one clock on an accelerated round trip and comparing it with an
unaccelerated clock before and after). Whereas, the so-called relativity of
simultaneity is merely the result of a convention about the way clocks are
synchronised in different frames.

It is worth remarking that, as already mentioned in section 2.2, the origin of
the term $-vx/c^{2}$ in (\ref{tprime}) was clearly understood by Poincar\'{e}
well before 1905. In a paper published in 1900 (Poincar\'{e} 1900), concerned
mainly with action and reaction in electrodynamics, Poincar\'{e} (who works to
lowest order in $v/c$) writes:\footnote{For a detailed analysis of this paper
by Poincar\'{e}, as well as for a reconstruction of Poincar\'{e}'s argument in
the cited passage, see the paper by Darrigol (1995).}

\begin{quotation}
`I assume that observers situated at different points set their watches with
the aid of light signals; that they try to correct these signals by the
transmission time, but that ignoring their translatory motion and therefore
believing that the signals are transmitted with equal speed in both
directions, they content themselves with crossing the observations, sending a
signal from A to B, then another from B to A. The local time $t%
\acute{}%
$ is the time shown by watches set in this way.

If then $V=1/\sqrt{K_{0}}$ is the speed of light, and $v$ the speed of
translation of the Earth which I assume parallel to the positive $x$-axis, one
will have:%
\[
t%
\acute{}%
=t-\frac{vx}{V^{2}}%
\]
' (Poincar\'{e} 1900: 483; translation by the author).
\end{quotation}

Poincar\'{e} understood that moving experimenters who assume that the speed of
light is still $c$ in all directions would adjust their clocks at different
points in space with settings that differ by the term $-vx/c^{2}$ (to lowest
order in $v/c$).

If instead distant clocks are synchronised by nonlocal means, then the speed
of light will be measured to be isotropic only in the preferred rest frame. In
quantum equilibrium, of course, such nonlocal signalling is impossible and the
true rest frame cannot be detected.

Note that, in the specific hidden-variables theory given by pilot-wave
dynamics, even leaving nonlocality aside, the natural kinematics of the theory
is arguably that of Aristotelian spacetime $E\times E^{3}$, with a preferred
state of rest (Valentini 1997). This is essentially because the dynamics is
first order in time, so that rest is the only reasonable definition of
`natural' or `unforced' motion. Pilot-wave theory then has a remarkable
internal logic: both the structure of the dynamics, and the operational
possibility of nonlocal signalling out of equilibrium, independently point to
the existence of a natural preferred state of rest.

\subsection{Curved Spacetime}

In the presence of gravitation, the above discussion may be extended to any
classical background spacetime possessing at least one global time function
$t$. This is hardly a restrictive requirement. For it is widely assumed that,
classically, any physical spacetime must be globally hyperbolic\footnote{See,
for example, Penrose (1979).} -- that is, must possess a Cauchy surface (a
spacelike slice on which initial data determine the entire spacetime) -- and
it is a theorem that any globally hyperbolic spacetime has topology
$\mathbb{R}\times\Sigma$ (where $\Sigma$ is a Cauchy surface) (Hawking and
Ellis 1973).

Consider, then, a curved spacetime that can be foliated (in general
non-uniquely) by spacelike hypersurfaces $\Sigma$ labelled by a global time
function $t$. The classical spacetime metric may then be written in the form%
\[
d\tau^{2}=\,^{(4)}g_{\mu\nu}dx^{\mu}dx^{\nu}=N^{2}dt^{2}-g_{ij}dx^{i}dx^{j}%
\]
where we have set the shift vector $N^{i}=0$, so that lines $x^{i}%
=\mathrm{const}.$ are normal to $\Sigma$. (This may always be done, as long as
the lines $x^{i}=\mathrm{const}.$ do not run into singularities.) The lapse
function $N(x^{i},t)$ measures the proper time lapse normal to $\Sigma$ per
unit of coordinate time $t$.

It may now be assumed that nonlocality acts instantaneously with respect to
one of these foliations, denoted $\Sigma(t)$. There is then a true slicing,
and spacetime is really the time evolution of the (absolute) 3-geometry
$\mathcal{G}(t)$ of $\Sigma(t)$, with metric $g_{ij}(x^{k},t)$ (Valentini
1992, 1996).

On this view, a small rod at time $t$ has proper length%
\[
dl=(g_{ij}dx^{i}dx^{j})^{1/2}%
\]
while a clock at rest in 3-space ticks a proper time%
\[
d\tau=N(x^{i},t)dt
\]
If a clock moves a spatial distance $dl$ in a time $dt$ it will tick a proper
time%
\[
d\tau^{2}=N^{2}dt^{2}-dl^{2}%
\]

Some remarks are in order.

First, there is an asymmetry here between space and time. It is assumed that
ordinary rods faithfully realise the true distance element $dl$ of space.
Whereas, we assume that ordinary clocks do not faithfully register true time
$t$; rather, their rate of ticking is affected by the local lapse field
$N(x^{i},t)$.

Second, note the difference from the case where gravity is absent. There we
saw that moving clocks are slowed down. The same effect occurs here, but in
addition, the rate of ticking of clocks is affected by their spatial location.
There is a field $N(x^{i},t)$ on 3-space which has a dynamical effect on the
rate of clocks even when they are at rest.

Third, this interpretation does not necessarily involve the introduction of an
independent field $N$ on 3-space. For this field could be determined by the
geometry of 3-space; $N$ could, for example, be a simple fixed function of the
3-metric $g_{ij}$ such as%
\[
N=g^{-1/2}\ \ \ \ \ \ (g\equiv\det g_{ij})
\]
(as in unimodular gravitation with $\det{}^{(4)}g_{\mu\nu}=1$ (Unruh 1989)).
Presumably, $N$ will be merely an effective field, emerging from some more
fundamental theory (possibly a quantum or subquantum theory of gravity). In
this way, there could be an underlying dynamical origin for the
phenomenological distortion of clock rates by the field $N$.

As in the flat case, one must be careful to distinguish between simultaneity
and synchronicity. Clocks located at different spatial points $x^{i}$ on the
same hypersurface (with label $t$) record simultaneous events, but the field
$N$ causes even stationary clocks to tick at different rates and lose their
synchrony. Thus, for example, let clocks at events $A_{1}$, $B_{1}$ at time
$t_{1}$ (on the preferred spacelike hypersurface $\Sigma_{1}$) move along
timelike lines to events $A_{2}$, $B_{2}$ at time $t_{2}>t_{1}$ (on the
preferred spacelike hypersurface $\Sigma_{2}$). Assume for simplicity that the
clocks remain at rest in space. Then each clock will tick a proper time%
\[
\Delta\tau=\int_{t_{1}}^{t_{2}}N(x^{i},t)dt
\]
where the integral is taken along the respective path. The lapse function $N$
will generally differ along the two paths $A_{1}$--$A_{2}$, $B_{1}$--$B_{2}$.
Thus, clocks synchronised at the simultaneous events $A_{1}$ and $B_{1}$
(using nonlocal signals) will no longer be synchronised at $A_{2}$ and $B_{2}
$, even though $A_{2}$ and $B_{2}$ are also simultaneous.

From a conventional perspective, this view will certainly seem eccentric, and
indeed it would be in the absence of any evidence for nonlocality. But if one
takes seriously Bell's deduction that nonlocal influences do occur in Nature,
and if one further accepts that our current inability to control these events
is merely a contingency of a particular distribution of hidden variables, and
bearing in mind that these effects occur at ordinary energies and macroscopic
distances, then the above view provides a consistent phenomenological means of
embedding such nonlocally-connected quantum events within the surrounding
approximately classical spacetime.

Again, one need not view the above construction as fundamental. A microscopic
theory of spacetime may well provide a very different picture at the
fundamental level. But if one accepts the existence of nonlocality, then it
seems natural that the above construction should emerge in some approximation.

In quantum equilibrium, of course, nonlocality and the true slicing cannot be
detected, as in the case of flat spacetime. Possibly, the observed
cosmological rest frame is a relic of early nonlocality -- arising from
quantum non-equilibrium in the early universe -- and coincides with true rest
(Valentini 1991b, 1992, 1996).

\section{Discussion and Conclusion}

We have presented a means of embedding quantum nonlocality within a background
classical spacetime (flat or curved), by introducing an absolute simultaneity
associated with a preferred foliation by spacelike hypersurfaces (where the
preferred foliation defines a preferred local state of rest). It should be
noted that this is unlikely to be the \textit{only} way of constructing such
an embedding. For as emphasised by Poincar\'{e}, the choice of geometry to be
used in physics is really dictated by convenience. There is no question of
proving that the most convenient choice is the only one possible, because one
may always adopt a different geometry by adding appropriate compensating
factors to the dynamics.

In his book \textit{Science and Hypothesis} (Poincar\'{e} 1902), Poincar\'{e}
illustrated this point in terms of an analogy with measuring rods affected by
thermal expansion. Consider, for example, metal rods on a heated flat metal
plate.\footnote{Poincar\'{e}'s example actually involved a 3-sphere within
which the temperature varies as a certain function of radius.} If the
temperature of the plate is non-uniform, and if all the rods have the same
expansion coefficient, then (assuming the rods reach thermal equilibrium
instantly) measurements within the surface using these rods will simulate the
geometry of a curved 2-surface -- that is, a non-Euclidean geometry. Creatures
living on such a surface could believe it to be curved, as long as all their
rods were affected by temperature in the same way. Equally, they could believe
their surface to be flat, with all rods being universally distorted (expanded
or contracted) by means of some agency acting upon them. There would be no way
of telling the difference. However, the creatures may well come to think that,
because the required distortions are the same for all rods, it is more
convenient to ascribe the distortions to the geometry of space itself; that
is, if the apparent geometry of the 2-surface is the same no matter which rods
are used, then one may as well define the apparent geometry to be the actual
geometry of space. As Poincar\'{e} put it:

\begin{quotation}
`Experiment .... tells us not what is the truest, but what is the most
convenient geometry' (Poincar\'{e} 1902).
\end{quotation}

The situation is no different in present-day physics. For example, instead of
interpreting general relativity in terms of a curved spacetime with metric
$g_{\mu\nu}$, it is possible to interpret it in terms of a Minkowski
spacetime, with flat metric $\eta_{\mu\nu}$, containing a field $h_{\mu\nu
}=g_{\mu\nu}-\eta_{\mu\nu}$ which distorts rods and clocks so as to give the
appearance of curved spacetime (Weinberg 1972). It cannot be proved that
spacetime is really curved; but, because the effects of the field $h_{\mu\nu}$
are universal -- the same for all rods and clocks -- it is more convenient to
regard those effects as purely kinematical, that is, as part of the geometry
with metric $g_{\mu\nu}$.

Similarly, classical special relativity may equally be interpreted in terms of
a preferred (yet unobservable) rest frame, where motion with respect to the
preferred frame has the dynamical effect of slowing clocks and contracting
rods. As emphasised for example by Bell (1987), this is an equivalent
formulation of the same physics. One may find it objectionable to have an
underlying preferred frame which can never be detected (classically), but
nevertheless this formulation of special-relativistic physics is consistent.
It often happens that the same physics can be formulated in equivalent,
empirically indistinguishable ways. Instead of insisting that non-standard
formulations are `wrong', it might be wiser to bear in mind that they might
prove useful in some situations, and that in the future, as new physics is
discovered, they might even turn out to be closer to the truth. The
preferred-frame interpretation of special relativity certainly comes into its
own in the face of quantum nonlocality.

These examples illustrate a general point. The division between kinematics and
dynamics cannot be determined uniquely. There is a `shifty split' between the
two. Yet, it is convenient to define the kinematics (or spacetime geometry) so
that it contains or summarises universal physical effects which are
independent of (for example) the mass and composition of bodies. For this
reason, universal symmetries such as Lorentz invariance are usually regarded
as part of the kinematics, so that spacetime is defined as locally
Minkowskian. However, with the discovery of new effects such as quantum
nonlocality, the most convenient choice of spacetime geometry may have to be
revised, as we have argued here.

From this `Poincar\'{e}an' point of view, it seems misguided to try to argue
that a certain kinematics -- with or without an absolute simultaneity --
\textit{must} be adopted. One can only propose a certain kinematics and argue
that it provides the simplest and most natural description of the
phenomena.\footnote{This is, in fact, arguably true not only regarding
spacetime geometry, but also regarding physical laws in general, since it is
always possible to write alternative formulations of the same physics.}

We claim, then, that the above construction, with an absolute simultaneity
(associated with a preferred foliation and a preferred local state of rest),
is the natural one given the known facts; and, we suggest that it should
emerge from a more fundamental theory in the limit of an approximately
classical spacetime background.

Alternatively, one might try to develop a theory of nonlocal interactions on
Minkowski spacetime. In itself, the mere fact of superluminal interaction is
not necessarily incompatible with fundamental Lorentz invariance. For example,
the interactions might be instantaneous in the centre-of-mass frame (a
manifestly Lorentz-invariant statement). But then one must somehow make sense
of backwards-in-time signals in other frames. This last question becomes
particularly poignant if one is willing to consider quantum non-equilibrium
and the associated practical signalling at a distance. Some workers, however,
maintain that backwards-in-time effects should be allowed, arguing that these
provide a loophole through which nonlocality may be avoided (Price
1996).\footnote{The derivation of Bell's inequality assumes that the initial
parameters $\lambda$ are unaffected by the future settings of the equipment.}
Attempts have been made to formulate a de Broglie-Bohm-type theory of particle
trajectories with fundamental Lorentz invariance, but it would appear that the
dynamics (and the quantum equilibrium distribution) must be defined on a
preferred spacelike slice, that is, in a preferred rest frame (Hardy 1992;
Berndl and Goldstein 1994; Berndl \textit{et al}. 1996). (See, however,
Dewdney and Horton (2002) for an attempt to avoid this problem.) A similar
result has been shown for any preferred local quantum observable (not
necessarily particle positions) (Myrvold 2002). In evaluating the advantages
and disadvantages of all these approaches, in our view, it ought to be
remembered that spacetime structure is not a metaphysical \textit{a priori}
background onto which dynamics is to be grafted at all costs; rather, it is as
subject to possible revision as dynamics itself.

It may well be that the issue of nonlocality vis \`{a} vis relativistic
spacetime will only be settled upon making further progress in physics. From
our perspective, for as long as we are confined to a state of statistical
equilibrium that hides the underlying nonlocality from direct view, it seems
probable that the argument will continue to be unresolved. On the other hand,
if quantum non-equilibrium were to be discovered and used in practice for
instantaneous signalling over remote distances, then in such circumstances it
seems likely that physicists would see the convenience of adopting a global
definition of absolute simultaneity.

\textbf{Acknowledgement.} I am grateful to Harvey Brown for allowing me to see
the manuscript of his book (Brown 2005) prior to publication, and for detailed
comments and correspondence regarding section 2 of this paper.

\ 

REFERENCES

\ 

Allen, R. E. (1997) "A statistical superfield and its observable
consequences," \textit{Int. J. Mod. Phys}. A 12, 2385--2412.

Amelino-Camelia, G. (2002) "Space-time quantum solves three experimental
paradoxes," \textit{Phys. Lett. B} 528, 181--187.

Aspect, A., J. Dalibard and G. Roger (1982) "Experimental test of Bell's
inequalities using time-varying analyzers," \textit{Phys. Rev. Lett.} 49, 1804--1807.

Bell, J. S. (1964) "On the Einstein-Podolsky-Rosen paradox," \textit{Physics}
1, 195--200.

Bell, J. S. (1986) "interview," in: \textit{The Ghost in the Atom}, eds. P. C.
W. Davies and J. R. Brown, Cambridge University Press, Cambridge.

Bell, J. S. (1987) \textit{Speakable and Unspeakable in Quantum Mechanics},
Cambridge University Press, Cambridge.

Berndl, K. and S. Goldstein (1994) "Comment on `Quantum mechanics, local
realistic theories, and Lorentz-invariant realistic theories'," \textit{Phys.
Rev. Lett.} 72, 780.

Berndl, K., D. D\"{u}rr, S. Goldstein and N. Zangh\`{\i} (1996) "Nonlocality,
Lorentz invariance, and Bohmian quantum theory," \textit{Phys. Rev. A} 53, 2062--2073.

Bohm, D. (1952a) "A suggested interpretation of the quantum theory in terms of
`hidden' variables. I," \textit{Phys. Rev.} 85, 166--179.

Bohm, D. (1952b) "A suggested interpretation of the quantum theory in terms of
`hidden' variables. II," \textit{Phys. Rev.} 85, 180--193.

Bohm, D. and B. J. Hiley (1984) "Measurement understood through the quantum
potential approach," \textit{Found. Phys.} 14, 255--274.

Born, M. (1962) \textit{Einstein's Theory of Relativity}, Dover, New York.

Brown, H. R. (2005) \textit{Physical Relativity: Space-Time Structure from a
Dynamical Perspective}, Oxford University Press, Oxford.

Chadha, S. and H. B. Nielsen (1983) "Lorentz invariance as a low energy
phenomenon," \textit{Nucl. Phys. B} 217, 125--144.

Coleman, S. and S. L. Glashow (1999) "High-energy tests of Lorentz
invariance," \textit{Phys. Rev. D} 59, 116008.

Colladay, D. and V. A. Kosteleck\'{y} (1998) "Lorentz-violating extension of
the standard model," \textit{Phys. Rev. D} 58, 116002.

Darrigol, O. (1995) "Henri Poincar\'{e}'s criticism of \textit{fin de
si\`{e}cle} electrodynamics," \textit{Stud. Hist. Phil. Mod. Phys.} 26, 1--44.

Darrigol, O. (1996) "The electrodynamic origins of relativity theory,"
\textit{Hist. Stud. Phys. Biol. Sci.} 26, 241--312.

de Broglie, L. (1928) "La nouvelle dynamique des quanta," in:
\textit{\'{E}lectrons et Photons: Rapports et Discussions du Cinqui\`{e}me
Conseil de Physique}, ed. J. Bordet, Gauthier-Villars, Paris, pp. 105--132.
[English translation: G. Bacciagaluppi and A. Valentini (forthcoming),
\textit{Electrons and Photons: the Proceedings of the Fifth Solvay Congress},
Cambridge University Press, Cambridge.]

Dewdney, C. and G. Horton (2002) "Relativistically invariant extension of the
de Broglie--Bohm theory of quantum mechanics," \textit{J. Phys. A }35, 10117--10127.

Einstein, A. (1905) "Zur Elektrodynamik bewegter K\"{o}rper," \textit{Ann. der
Phys. }17, 891--921.

Einstein, A. (1949a) "Remarks concerning the essays brought together in this
co-operative volume," in: \textit{Albert Einstein: Philosopher-Scientist}, ed.
P. A. Schilpp, Open Court, Illinois, p. 678.

Einstein, A. (1949b) "Autobiographical notes," in: \textit{Albert Einstein:
Philosopher-Scientist}, ed. P. A. Schilpp, Open Court, Illinois.

Einstein, A. (1993) \textit{The Collected Papers of Albert Einstein}, volume
3, Princeton University Press, Princeton. [Original paper: A. Einstein (1910)
"Le principe de relativit\'{e} et ses cons\'{e}quences dans la physique
moderne," \textit{Archives des Sciences Physiques et Naturelles} 29, 5--28.]

Einstein, A., B. Podolsky and N. Rosen (1935) "Can quantum-mechanical
description of physical reality be considered complete?," \textit{Phys. Rev.
}47, 777--780.

Ernst, A. and J.-P. Hsu (2001), in: \textit{Lorentz and Poincar\'{e}
Invariance: 100 Years of Relativity}, eds. J.-P. Hsu and Y.-Z. Zhang, World
Scientific, Singapore, pp. 4--24.

Granek, G. (2000) "Poincar\'{e}'s contributions to relativistic dynamics,"
\textit{Stud. Hist. Phil. Mod. Phys.} 31, 15--48.

Hardy, L. (1992) "Quantum mechanics, local realistic theories, and
Lorentz-invariant realistic theories," \textit{Phys. Rev. Lett.} 68, 2981--2984.

Hawking, S. W. and G. F. R. Ellis (1973) \textit{The Large Scale Structure of
Space-Time}, Cambridge University Press, Cambridge.

Jacobson, T. and D. Mattingly (2001) "Gravity with a dynamical preferred
frame," \textit{Phys. Rev. D} 64, 024028.

Janssen, M. and J. Stachel (2004), "The optics and electrodynamics of moving
bodies," preprint 265, Max Planck Institute for the History of Science.

Kilmister, C. W. (1970) \textit{Special Theory of Relativity}, Pergamon, New York.

Kosteleck\'{y}, V. A. (2002) (ed.) \textit{Proceedings of the Second Meeting
on CPT and Lorentz Symmetry}, World Scientific, Singapore.

Mavromatos, N. E. (2004) "CPT violation and decoherence in quantum gravity," gr-qc/0407005.

Miller, A. I. (1973) "A study of Henri Poincar\'{e}'s `Sur la dynamique de
l'\'{e}lectron'," \textit{Arch. Hist. Exact Sci.} 10, 207--328 (1973).
[Reprinted (1986), in: A. I. Miller, \textit{Frontiers of Physics:
1900--1911}, Birkh\"{a}user, Boston, pp. 29--150.]

Minkowski, H. (1952) "Space and time," in: \textit{The Principle of
Relativity}, Dover, New York, pp. 75--91.

Moffat, J. W. (2003) "Spontaneous violation of Lorentz invariance and
ultra-high energy cosmic rays," \textit{Int. J. Mod. Phys. D} 12, 1279--1287.

Nielsen, H. B. and M. Ninomiya (1978) "$\beta$-Function in a non-covariant
Yang-Mills theory," \textit{Nucl. Phys. B} 141, 153--177.

Pais, A. (1982) \textit{Subtle is the Lord: the Science and the Life of Albert
Einstein}, Oxford University Press, Oxford.

Pauli, W. (1958) \textit{Theory of Relativity}, Pergamon Press, London.

Penrose, R. (1979) "Singularities and time-asymmetry," in: \textit{General
Relativity: an Einstein Centenary Survey}, eds. S. W. Hawking and W. Israel,
Cambridge University Press, Cambridge, pp. 581--638.

Poincar\'{e}, H. (1900) "La th\'{e}orie de Lorentz et le principe de
r\'{e}action," \textit{Archives n\'{e}erlandaises des Sciences exactes et
naturelles}, second series, vol. 5, pp. 252--278. [Reprinted (1954), in:
\textit{Oeuvres de Henri Poincar\'{e}}, volume IX, Gauthier-Villars, Paris,
pp. 464--488.]

Poincar\'{e}, H. (1902) \textit{La Science et l'Hypoth\`{e}se}, Flammarion,
Paris. [English translation (2001): \textit{The Value of Science: Essential
Writings of Henri Poincar\'{e}}, Modern Library, New York, p. 59.]

Poincar\'{e}, H. (1905a) \textit{La Valeur de la Science}, Flammarion, Paris.
[English translation (2001): \textit{The Value of Science: Essential Writings
of Henri Poincar\'{e}}, Modern Library, New York, p. 222.]

Poincar\'{e}, H. (1905b) "Sur la dynamique de l'\'{e}lectron," \textit{C. R.
Ac. Sci. Paris} 140, 1504--1508. [Reprinted (1954), in: \textit{Oeuvres de
Henri Poincar\'{e}}, volume IX, Gauthier-Villars, Paris, pp. 489--493.]

Poincar\'{e}, H. (1906) "Sur la dynamique de l'\'{e}lectron," \textit{Rend.
Circ. Matem. Palermo} 21, 129--176. [Reprinted (1954), in: \textit{Oeuvres de
Henri Poincar\'{e}}, volume IX, Gauthier-Villars, Paris, pp. 494--550.]

Poincar\'{e}, H. (1910) "La m\'{e}canique nouvelle," in: \textit{Sechs
Vortr\"{a}ge \"{u}ber Ausgew\"{a}hlte Gegenst\"{a}nde aus der Reinen
Mathematik und Mathematischen Physik}, Teubner, Leipzig, pp. 51--58.

Popper, K. R. (1982) \textit{Quantum Theory and the Schism in Physics}, Unwin
Hyman, London.

Price, H. (1996) \textit{Time's Arrow and Archimedes' Point}, Oxford
University Press, Oxford.

Schwartz, H. M. (1971) "Poincar\'{e}'s Rendiconti paper on relativity. Part
I," \textit{Am. J. Phys.} 39, 1287--1294.

Schwartz, H. M. (1972) "Poincar\'{e}'s Rendiconti paper on relativity. Part
II," \textit{Am. J. Phys.} 40, 862--872.

Unruh, W. G. (1989) "Unimodular theory of canonical quantum gravity,"
\textit{Phys. Rev. D} 40, 1048--1052.

Valentini, A. (1991a) "Signal-locality, uncertainty, and the subquantum
\textit{H}-theorem. I," \textit{Phys. Lett. A} 156, 5--11.

Valentini, A. (1991b) "Signal-locality, uncertainty, and the subquantum
\textit{H}-theorem. II," \textit{Phys. Lett. A} 158, 1--8.

Valentini, A. (1992) \textit{On the Pilot-Wave Theory of Classical, Quantum
and Subquantum Physics}, PhD thesis, International School for Advanced
Studies, Trieste, Italy.

Valentini, A. (1996) "Pilot-wave theory of fields, gravitation and cosmology,"
in: \textit{Bohmian Mechanics and Quantum Theory: an Appraisal}, eds. J. T.
Cushing, A. Fine, and S. Goldstein, Kluwer, Dordrecht, pp. 45--66.

Valentini, A. (1997) "On Galilean and Lorentz invariance in pilot-wave
dynamics," \textit{Phys. Lett. A} 228, 215--222.

Valentini, A. (2001) "Hidden variables, statistical mechanics and the early
universe," in: \textit{Chance in Physics: Foundations and Perspectives}, eds.
J. Bricmont \textit{et al}., Springer, Berlin, pp. 165--181.

Valentini, A. (2002a) "Signal-locality in hidden-variables
theories,"\textit{\ Phys. Lett. A} 297, 273--278.

Valentini, A. (2002b) "Signal-locality and subquantum information in
deterministic hidden-variables theories," in: \textit{Non-locality and
Modality}, eds. T. Placek and J. Butterfield, Kluwer, Dordrecht, pp. 81--103.

Valentini, A. (2002c) "Subquantum information and computation,"
\textit{Pramana -- J. Phys.} 59, 269--277.

Valentini, A. (2004a) "Black holes, information loss, and hidden variables," hep-th/0407032.

Valentini, A. (2004b) "Extreme test of quantum theory with black holes," astro-ph/0412503.

Valentini, A. and H. Westman (2005) "Dynamical origin of quantum
probabilities," \textit{Proc. Roy. Soc. A} 461, 253--272.

Voigt, W. (1887) "\"{U}ber das Dopplersche Prinzip," \textit{Nachr. Ges. Wiss.
G\"{o}ttingen} 41. [English translation, in: Ernst and Hsu (2001: 10--19).]

Weinberg, S. (1972) \textit{Gravitation and Cosmology}, Wiley, New York.

Weinberg, S. (1995) \textit{The Quantum Theory of Fields, vol. I,
Foundations}, Cambridge University Press, Cambridge.

Zahar, E. (1989) \textit{Einstein's Revolution: a Study in Heuristic,} Open
Court, La Salle.
\end{document}